# Spectroscopic evidence that the extreme properties of IRAS F 10214+4724 are due to gravitational lensing

Stephen Serjeant[1], Mark Lacy[1], Steve Rawlings[1], Lindsay J. King[1] & D. L. Clements[2]

[1] *Astrophysics, Department of Physics, Keble Road, Oxford, OX1 3RH*
[2] *European Southern Observatory, Karl-Schwarzschild-Strasse 2, D-85748, Garching-bei-München, Germany*

14 June 1995

**ABSTRACT**
The extreme bolometric luminosity of IRAS F 10214+4724, and in particular the huge mass in molecular gas inferred from the CO line fluxes have led to suggestions that this is a giant galaxy in the process of formation. An arc-like structure and the closeness of a second object suggest, however, that gravitational lensing may be responsible for its anomalously high luminosity and mass. In this paper, we use an optical spectrum taken in conditions of 0.7-arcsec seeing to provide further evidence that F 10214+4724 is a gravitationally lensed system. We measure tentative redshifts of 0.896 and 0.899 for galaxies projected $\approx 1$ and $\approx 3$ arcsec from IRAS F 10214+4724 respectively. Identifying the former as the lensing galaxy we obtain a mass:light ratio consistent with those derived for other lenses, and find that its luminosity is consistent with the velocity dispersion deduced from the radius of the Einstein ring. If lensed, our models suggest magnification by a factor $\sim 10$, and hence an intrinsic bolometric luminosity for F 10214+4724 similar to local ULIRGs.

## 1 INTRODUCTION

Although IRAS F 10214+4724 lies below the completeness limit of the IRAS faint source catalogue, its redshift of $z = 2.286$ and its far-infrared and sub-mm fluxes imply an enormous bolometric luminosity of $\sim 3 \times 10^{14} L_\odot$ * (e.g. Rowan-Robinson et al. 1991). Indeed, IRAS F 10214+4724 is more than an order of magnitude more luminous than any local ultraluminous infrared galaxy (ULIRG) (Rowan-Robinson et al. 1993). Moreover, the $H_2$ mass inferred from its CO luminosity exceeds that of any other known galaxy by a factor of $\approx 3$ (Solomon, Downes & Radford 1992). This huge mass of molecular gas, unique in being comparable to the total mass of stars and gas in the Milky Way, has been interpreted as a primaeval giant galaxy.

Recent sub-arcsecond near-infrared images from the Keck telescope (Matthews et al. 1994; hereafter M94) showed the IRAS galaxy has an arc-like structure centred on a nearby ($\approx 1$ arcsec distant) object. M94 attributed this structure to the tidal forces induced in a galaxy-galaxy interaction. On the basis of the magnitudes and colours of both the close neighbour and a further object $\approx 3$ arcsec from F10214+4724, however, Elston et al. (1994) suggested that F 10214+472 is being lensed by a galaxy in a small group. If gravitationally lensed, F 10214+4724 may not intrinsically have such extreme properties: a magnification of $\gtrsim 10$ would make it a fairly typical member of the ULIRG class.

In this paper we present optical spectra of the two

---

* We assume a Hubble constant $H_0 = 50$ kms$^{-1}$Mpc$^{-1}$, a density parameter $\Omega_0 = 1$ and a cosmological constant of zero

objects close to F 10214+4724 which we use to provide new evidence in favour of the lensing hypothesis. The data obtained simultaneously on F 10214+4724 itself will be discussed elsewhere (Serjeant, Rawlings & Lacy in preparation).

## 2  DATA ACQUISITION AND ANALYSIS

Using the ISIS spectrograph on the William Herschel Telescope (WHT), we observed F 10214+4724 on the nights 1995 January 28 and 1995 January 30, taking advantage of good seeing conditions to complete our own WHT Service programme. We used a 2-arcsec wide long slit aligned at PA 22 to include light from the two objects close to the VLA position, RA(1950) $10^h 21^m 31^s.14$, Dec.(1950) $+47°24'22.9''$ (Rowan-Robinson et al. 1991). We offset from a star at RA(1950) $10^h 21^m 17^s.38$, Dec.(1950) $+47°22'59.7''$ measured using the GASP system at STScI. We used both arms of ISIS, splitting the beam with the 540 nm dichroic. The red arm grating was the R158R, and the detector was the TEK2 CCD; in the blue arm the grating was the R158B, and the detector was the TEK1 CCD. Over the two nights we exposed for $4 \times 900s + 1 \times 1800s$ in the red and $3 \times 1800s$ in the blue. A nearby star of spectral type F [RA(1950) $10^h 23^m 19s.32$, Dec.(1950) $+46°25'29.3''$] was observed to enable accurate removal of atmospheric absorption. The standard stars HD19445 and HZ44 were used for absolute flux calibration. Despite possible thin cirrus on both nights, the spectrophotometric calibration is accurate to $\pm 10$ per cent. The seeing was $\approx 0.7$ arcsec on both nights. The data were analysed using standard IRAF routines.

Spectra of sources 2 and 3 of M94 were then extracted from the final 2D spectrum. (Source 2 is the galaxy $\approx 1$ arcsec north of the arc-like image of source 1, and the likely lens candidate; source 3 is $\approx 2$ arcsec to the northeast of source 2.) The spectrum of source 2 was corrected for contamination from the nearby IRAS galaxy (source 1) by subtracting its spectrum, scaled by the ratio of the CIII]190.9 line in the spectrum of source 1 to that in source 2. These spectra are presented in Fig. 1. Apart from a marginally significant ($\sim 2 - 3\sigma$) emission feature at 658.5 nm in the spectrum of source 2 there are no emission lines in either spectrum. Both, however, show very similar spectral shapes with fairly pronounced spectral breaks around 760 nm. Although this is close to the atmospheric A band, we are confident that the breaks are not artefacts of imperfect atmospheric correction because, as illustrated in Fig. 1, these would manifest themselves as relatively narrow spikes.

The spectral breaks can be identified with the '4000-Å break' familiar from the spectra of galaxies. By cross-correlating our spectra with an old galaxy model (a 1-Gyr burst aged by $10^8$ yr) from Rocca-Volmerange & Guiderdoni (1988) we obtained red-shift estimates of 0.896 and 0.899 for sources 2 and 3 respectively; although these are tentative due to the low signal-to-noise of the spectra, the spectral energy distribution (SED) of source 3 (including additional near-infrared data from M94) provides a good match to the model (Fig. 1). Source 2 is also a fair match if our optical spectrum is scaled by a factor of $\approx 1.8$. A large scaling correction is reasonable because there will be some unknown aperture correction due to the non-colinearity of F 10214+4724, source 2 and source 3, and because source 2 is heavily contaminated by light from source 1.

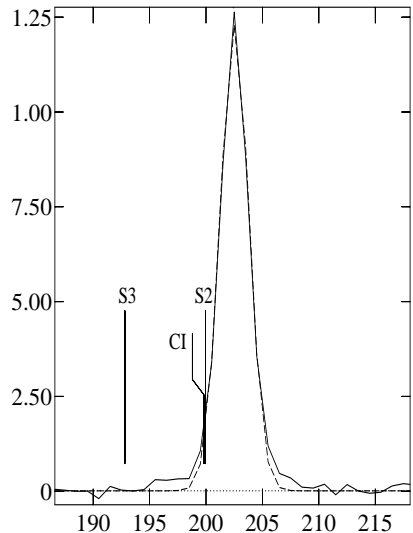

**Figure 2.** A profile cut across the continuum-subtracted CIV 154.9 line of F 10214+4724: the abscissa is labelled in pixels, each pixel being 0.35 arcsec; the ordinate is in arbitrary units. The solid line represents the data (averaged over 5 nm); the dashed line is a Gaussian fit. We estimated the continuum, by averaging the profile over 30 nm blueward of the emission line. The positions of sources 2 (S2) and 3 (S3), and the expected position of the counterimage (CI) are marked.

We searched our spectrum for a counterimage of source 1 by looking for emission features at the wavelengths of the highest signal-to-noise lines in the source 1 spectrum. After subtraction of the contribution of the continuum from the lens, no evidence for a counterimage was seen around the highest signal-to-noise line (CIV 154.9) to a $3\sigma$ limit of image to counterimage ratio of 70:1 (Fig. 2). The intrinsic ratio may be significantly smaller than this for two reasons: first, there will be an aperture correction, the magnitude of which will depend on the exact position of the counterimage; second, the ray path of the counterimage passes within a few kpc of the centre of the lensing galaxy, so extinction in this galaxy would increase the observed image:counterimage ratio. Al-though similarly large ratios ($\gtrsim$ 100) are suggested by the results of narrow-band near-infrared imaging (M94), the extinction caused by the lensing galaxy could remain high even at rest-frame wavelengths of $\sim 1\mu$m. A dusty lens model of this type seems necessary to explain the properties of at least two other known lens systems (e.g. Larkin et al. 1994 and references therein).

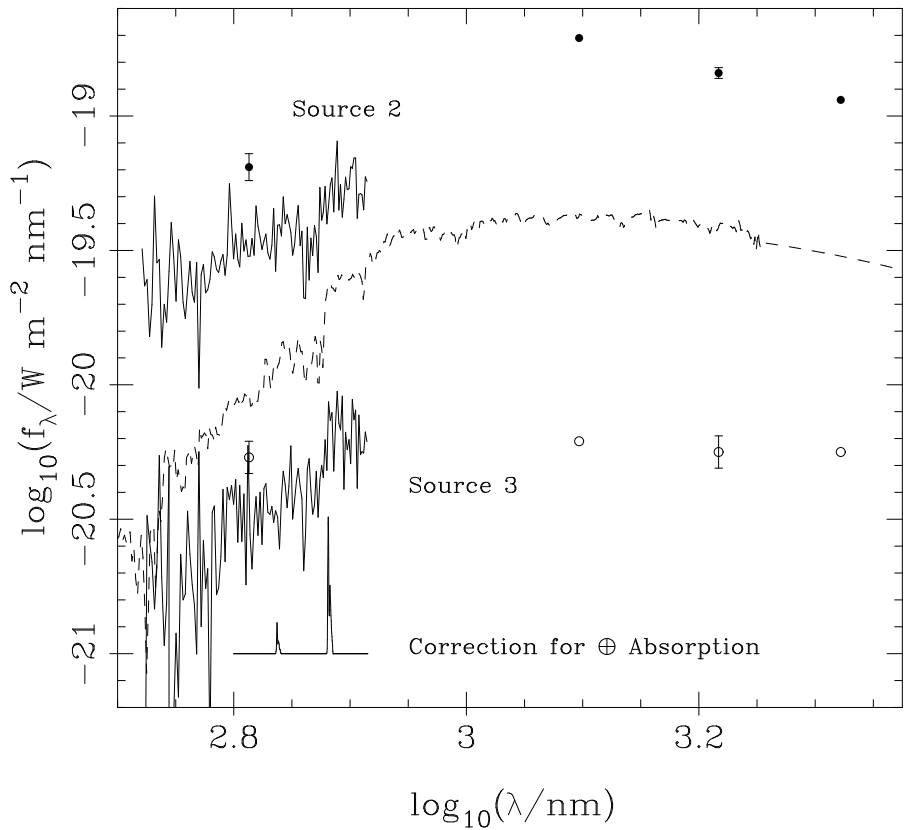

**Figure 1.** The spectral energy distributions of sources 2 and 3. The dashed line is the SED of the model galaxy discussed in the text, redshifted to 0.896. Above, displaced by 0.5 in $\lg(f_\lambda)$ is the SED of source 2, with our spectrum shown as the solid line and the broad-band measuremants of M94 and Elston et al. (1994) as filled circles. Below, the SED of source 3 is plotted, displaced by -0.5 in $\lg(f_\lambda)$ relative to the model, with broad-band points as open circles. On the same logarithmic scale we also plot the multiplicative correction made to the data to account for atmospheric absorption in the B and A bands.

## 3 A LENSING MODEL FOR F10214+4724

Without a high-resolution optical image, the parameters of the lens model cannot be uniquely constrained. Nevertheless, we can state some general features. First, the evident weakness of the counterimage suggests image formation close to a cusp in the source plane (Narayan & Wallington 1992), with consequent high magnification $\sim 10$. Second, we can obtain a direct estimate of the Einstein ring size $\theta_E$ of $\approx 0.8$-1.5 arcsec from the near-infrared structure (M94).

We used the GLENS task in the $\mathcal{AIPS}$ software package to model the system, using a Blandford-Kochanek lens potential (Blandford & Kochanek 1987). We found lens models in qualitative agreement with the image structure and limits on the image:counterimage ratio; typically the models have magnifications $\gtrsim 10$ and ellipticities ($\epsilon = (a^2 - b^2)/(a^2 + b^2)$ where $a$ and $b$ are the semi-major and semi-minor axes of the galaxy potential) between 0.1 and 0.2. The core radius was found to have little impact on the lens models for values $\lesssim 0.1\theta_E$, safely including all reasonable values for an elliptical galaxy. An example is shown in Fig. 3; this model has $\theta_E = 0.93$ arcsec, an ellipticity of 0.2 and zero core radius. This model gave a source magnification of 11 and an image:counterimage ratio of 40:1 (which could increase still further in a harder potential; Blandford & Kochanek 1987). A further class of lens models which could be made to reproduce our limit on the image to counterimage ratio have the source straddling a naked cusp; this may be produced by a galaxy potential with a strong shear component which could, for example, be enhanced by source 3 (Schneider, Ehlers & Falco 1992). Assuming a lens redshift of 0.896 and a value of $\theta_E = 0.93$ arcsec, we obtain an enclosed mass of $4.4 \times 10^{11} M_\odot$ assuming a negligible contribution to the lensing potential from source 3. Alternatively, using an isothermal sphere model (Turner, Ostriker & Gott 1984), our value of $\theta_E$ implies a (one-dimensional) velocity dispersion of 270 kms$^{-1}$.

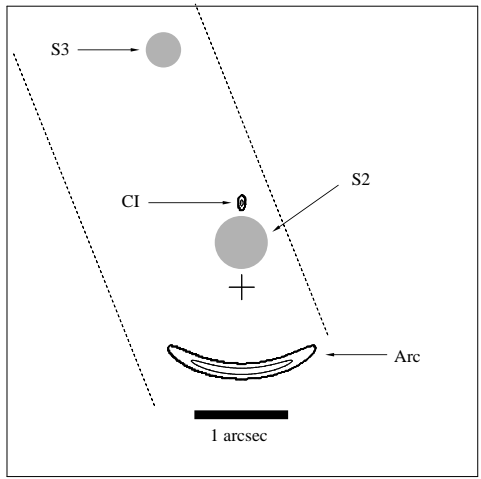

**Figure 3.** The lens model discussed in the text. The source position (in the absence of the lens) is marked with a cross. The source lies near a cusp in the source plane. The box is a square of side 5 $\theta_E$. The approximate locations of the sides of the slit are indicated by the dashed lines, and the positions of source 2 (S2) and 3 (S3) indicated by shading. The predicted positions of the arc and the counterimage (CI) are shown as contours.

These values of the mass can be used, in conjunction with the measured magnitudes of source 2 to estimate the mass:light ratio of the lens. We obtain a blue mass:light ratio of 6 if we take the M94 $K_s$ magnitude of 17.6, using the model galaxy to obtain K corrections and the enclosed mass derived above; this calculation assumes all the measured galaxy light is within the Einstein radius. The magnitude predicted from the velocity dispersion *via* the Faber-Jackson relation (Faber & Jackson 1976) is $K_s = 17.7$. The good agreement between these methods, and the mass:light ratios obtained from other lenses (e.g. Langston et al. 1990) adds further support to the lensing scenario.

## 4  DISCUSSION

Although the lens redshift is far from totally secure, it seems highly unlikely that source 2, the lens candidate, is at the redshift of F 10214+4724. If it were its luminosity would have to be as extreme as F 10214+4724, and the similarity of its spectral energy distribution to source 3 would have to be either a remarkable coincidence, or evidence for yet another extreme $z = 2.29$ galaxy. From the spectral energy distributions of sources 2 and 3 it seems far more likely that both are galaxies at $z \approx 0.9$. Assuming then that source 2 is foreground, our lensing model predicts a very reasonable $M/L$ ratio.

The colour gradients in source 1 have been used as an argument against lensing (e.g. M94). We believe this argument is weak since local ULIRGs have colour gradients (e.g. Carico et al. 1988), and the colour variations in F 10214+4724 are inevitably intrinsic if the source is not lensed. Another possible argument against lensing is the apparent lack of a counterimage. This can easily be explained by dust within the lensing galaxy as argued in other confirmed lens systems.

Gravitational lensing has a significant impact on the interpretation of F 10214+4724. Magnification by a factor of $\sim 10$ implies an intrinsic luminosity within the range spanned by local ULIRGs and IRAS-detected quasars. The dynamical mass inferred from the CO line width can now be reconciled with the mass inferred from the CO luminosity without invoking a face-on geometry for the molecular material (Solomon et al. 1992). F 10214+4724 also makes an interesting comparison with the only other high-$z$ object detected in CO, the 'Cloverleaf Quasar' (Barvainis et al. 1994); in both cases, the intrinsic CO emission is likely to be greatly magnified. Therefore, it may be that neither object is intrinsically much more luminous in molecular lines than local starburst systems. In short, if lensed, F 10214+4724 is no longer a unique object, and no stronger a protogalaxy candidate than starburst galaxies or quasars in the relatively local Universe.

There may also be more general implications for the study of star formation in high-redshift galaxies and its link with galaxy formation. It is possible, for example, that gravitational lensing may introduce significant biases in the next generation of surveys in the far-infrared (ISO) and sub-mm (SCUBA). Selection on the basis of extended radio emission should reduce such biases substantially. Unlike the lensed system F 10214+4724, the detection of rest-frame far-infrared emission from the $z = 3.8$ radiogalaxy 4C41.17 (Dunlop et al. 1994) has no local analogue, and may be the first example of the protogalaxy phase in a high-redshift system.

## 5  ADDENDUM

Subsequent to the submission of this paper, two further groups have suggested a gravitational lens interpretation for this system on the basis of the optical and near-infrared images (Broadhurst & Lehár 1995; Graham & Liu 1995). As well as confirming the arc-like structure of the near-infrared continuum first reported by M94, they both suggest tentative identi-fication of a counterimage close to the position predicted in Fig. 3 (and by their own lens modelling). To explain the lack of any emission lines in the candidate counterimage (Section 2; Soifer et al. 1992) suggests differential magnification between the emission line region and a more extended source of the near-infrared light (Broadhurst & Lehár 1995). Another recent paper (Soifer et al. 1995) attempted to obtain a spectrum of source 2 but no features were reported, probably because of their poorer seeing, and the lack of source 3 as a comparison.

## ACKNOWLEDGEMENTS

We thank Carlos Martin for assisting the observations, and the anonymous referee for useful comments. The WHT is operated on the island of La Palma by the Royal Greenwich Observatory in the Spanish Observatorio del Roque de los Muchachos of the Instituto de Astrofisica de Canarias. We thank Steve Eales for performing the GASP astrometry of the F10214 + 4724 field, and for (unwittingly) contributing observing time to the project.